\documentclass[twoside,reqno]{mbook}
\usepackage{epsfig}
\usepackage{amssymb,amsmath}
\usepackage{times}
\usepackage{color}
\newcommand{\SU}[1]{\ensuremath{\mathrm{SU}( #1 )}}
\newcommand{\Un}[1]{\ensuremath{\mathrm{U}( #1 )}}
\newcommand{\SO}[1]{\ensuremath{\mathrm{SO}( #1 )}}

\newcommand{\SpR}[1]{\ensuremath{\mathrm{Sp}( #1,\mathbb{R} )}}
\newcommand{\su}[1]{\ensuremath{\mathfrak{su}( #1 )}}

\newcommand{\so}[1]{\ensuremath{\mathfrak{so}( #1 )}}

\newcommand{\spR}[1]{\ensuremath{\mathfrak{sp}( #1, \mathbb{R} )}}
\newcommand{\ket}[1]{\ensuremath{\left| #1 \right\rangle}}

\newcommand{\ho}{\ensuremath{\hbar\Omega}}
\newcommand{\ph}[1]{\ensuremath{#1}p-\ensuremath{#1}h}
\begin{document}

\sloppy \raggedbottom

 \setcounter{page}{1}

\newpage
\setcounter{figure}{0}
\setcounter{equation}{0}
\setcounter{footnote}{0}
\setcounter{table}{0}
\setcounter{section}{0}



\title{Light Nuclei in the Framework of the Symplectic No-core Shell Model}

\runningheads{Light Nuclei in the Framework of the Symplectic No-core Shell
Model}{J.P.~Draayer, T.~Dytrych, K.D.~Sviratcheva, C.~Bahri, and J.P.~Vary}

\begin{start}
\author{Jerry P. Draayer}{1}, \coauthor{Tom\'a\v{s} Dytrych}{1},
\coauthor{Kristina D. Sviratcheva}{1}, \coauthor{Chairul Bahri}{1},
\coauthor{James P. Vary}{2}

\address{Department of Physics and Astronomy, Louisiana State
University, Baton
Rouge, LA 70803, USA}{1}

\address{Department of Physics and Astronomy, Iowa State
University, Ames, IA 50011,
USA,\\
Lawrence Livermore National Laboratory, L-414, 7000 East
Avenue, Livermore,
California, 94551, USA, and\\
Stanford Linear Accelerator Center, MS81, 2575 Sand Hill
Road, Menlo Park,
California, 94025, USA}{2}

\begin{Abstract}
A symplectic no-core shell model (Sp-NCSM) is constructed with the goal
of extending the {\it ab-initio} NCSM to include strongly deformed
higher-oscillator-shell configurations and to reach heavier nuclei that
cannot be studied currently because the spaces encountered are too large
to handle, even with the best of modern-day computers. This goal is
achieved by integrating two powerful concepts: the {\it ab-initio} NCSM
with that of the $\SpR{3} \supset \SU{3}$ group-theoretical approach. The
NCSM uses modern realistic nuclear interactions in model spaces that
consists of many-body configurations up to a given number of
\ho~ excitations together with modern high-performance parallel
computing techniques. The symplectic theory extends this picture by
recognizing that when deformed configurations dominate, which they often
do, the model space can be better selected so less relevant low-lying
\ho~ configurations yield to more relevant high-lying
\ho~ configurations, ones that respect a near symplectic
symmetry found in the Hamiltonian. Results from an application of the
Sp-NCSM to light nuclei are compared with those for the NCSM and with
experiment.
\end{Abstract}
\end{start}

\section[]{Introduction}
The concept of an {\it ab initio} no-core shell-model (NCSM) \cite{NCSM}, which
yields a  good description of
the low-lying states in few-nucleon systems as
well as in more complex nuclei \cite{NCSM,NavratilBK98_00}, has taken center stage in 
the development of microscopic tools for studying the structure of  atomic nuclei. ÊThe
architecture for the NCSM capitalizes on  computational efficiencies that can be
realized when many-particle  Slater determinant basis states are mapped onto an
integer bit string  representation of that state on a computer. In addition, in the
framework of the NCSM one can employ modern realistic interactions that reflect on the
essence of the strong interaction. Recently developed realistic $NN$ potentials
include $J$-matrix inverse scattering potentials \cite{ShirokovMZVW04}, high-precision
$NN$ potentials derived from meson exchange theory \cite{MachleidtSS96M01} and nuclear
two- and many-body forces based on chiral effective field theory \cite{EntemM03}.

The symplectic no-core  shell-model (Sp-NCSM) \cite{DytrychSBDV06} amplifies on this
concept by recognizing that  deformed configurations often dominate and these, while
typically described by only few collective \SpR{3} basis states, correspond to a 
special linear combination of a large number of NCSM basis states. Hence, the
effective size  of the model space can be significantly reduced and constrained to 
respect a near symplectic symmetry, that within a 0\ho~ space reduces  to SU(3), of
the model Hamiltonian. In this way, the Sp-NCSM will allow one to account for even
higher $\hbar\Omega$ configurations required to realize experimentally measured B(E2)
values without an effective charge, and especially highly deformed spatial
configurations required to reproduce
$\alpha$-cluster modes in heavier nuclei.

As a `proof-of-principle' study,  results for no-core and symplectic no-core
calculations up to 6\ho~ are compared for two nuclei, namely, the deformed $^{12}$C and
the closed-shell $^{16}$O.  The analysis of the results shows that the
$0^{+}_{gs}$ and the lowest $2^{+}$ and $4^{+}$ states in $^{12}$C as well as
the $0^{+}_{gs}$ in  $^{16}$O, which are derived in the framework of the NCSM with the
JISP16 realistic interaction \cite{ShirokovMZVW04} and are well converged,
reflect the presence of an underlying symplectic \spR{3} algebraic structure
\footnote{We use lowercase (capital) letters for algebras (groups).}.  This is
achieved through the projection of realistic NCSM eigenstates onto 
\SpR{3}-symmetric basis states of the symplectic shell model.

The symplectic shell model \cite{Sp3R1,Sp3R2} is a multiple oscillator shell
generalization of Elliott's \SU{3} model and as well, a microscopic realization of the
successful Bohr-Mottelson collective model. Symplectic
algebraic approaches have achieved a very good reproduction of
low-lying energies in $^{12}$C using phenomenological interactions \cite{EscherL02} or
truncated symplectic basis with simplistic (semi-) microscopic interactions
\cite{ArickxBD82,AvanciniP93}. Here, we establish, for the
first time, the dominance of the symplectic \SpR{3} symmetry in nuclei as unveiled
through {\it ab initio} calculations of the NCSM type with realistic interactions.
This in turn opens up a new and exciting possibility
for representing significant high-$\hbar\Omega$ collective modes 
by extending the NCSM basis space beyond its current limits 
through \SpR{3} basis states, 
which yields a dramatically smaller basis space to achieve convergence
of higher-lying collective modes.
In this regard, it may be interesting to understand the importance of
a larger model space beyond the $6\hbar \Omega $ limit and its role in
shaping other low-lying states in $^{12}$C and $^{16}$O such as the second
$0^+$, which is likely to reflect a cluster-like behavior (e.g., see
\cite{FunakiTHSR03}). This task, albeit challenging, is feasible for the no-core shell
model with the symplectic
\SpR{3} extension.

\section{Symplectic Shell Model}

The symplectic shell model is based on the noncompact symplectic
\spR{3} algebra
that with its subalgebraic structure unveils the underlying physics
of a microscopic
description of collective modes in nuclei \cite{Sp3R1,Sp3R2}. The latter follows from
the fact that the
mass quadrupole and monopole moment operators, the many-particle
kinetic energy, the
angular and vibrational momenta are all elements of the $\spR{3}\supset \su{3}
\supset\so{3}$  algebraic structure. Hence, collective states of a nucleus with
well-developed quadrupole and monopole vibrations as well as
collective rotations are
described naturally in terms of irreducible representations (irreps)
of \SpR{3}.
Furthermore, the elements of the \spR{3} algebra are constructed as
bilinear products
in the harmonic oscillator (HO) raising and lowering 
operators that in turn are expressed
through particle
coordinates and linear momenta. This means the basis states of a
\SpR{3} irrep can be expanded in a HO ($m$-scheme) basis,
the same basis used in the NCSM, thereby facilitating symmetry identification. 

The symplectic basis states are labeled (in standard notation \cite{Sp3R1,Sp3R2})
according to the reduction chain
\begin{equation}
\begin{tabular}{ccccc}
$\SpR{3}$ & ~$\supset $ & $~~~\Un{3}$  & $\supset$  &    ~~~$\SO{3}$ \\
~~$\Gamma_\sigma\;\;$& $\Gamma_n\rho$ & ~~~$\Gamma_\omega$ &  $\kappa$  & ~~$L$
\end{tabular}
\end{equation}
and are constructed by acting with polynomials $\mathcal{P}$ in the
symplectic raising
operator, $A^{(2\,0)}$, on a set of basis states of the symplectic
bandhead, $\ket{\Gamma_{\sigma}}$, which is a \SpR{3} lowest-weight
state\footnote{A \SpR{3} lowest-weight
state, $\ket{\Gamma_{\sigma}}$, is defined as $A^{(0\,2)}|\Gamma_\sigma\rangle = 0$,
where the symplectic lowering operator $A^{(0\,2)}$ is the adjoint of $A^{(2\,0)}$.},
\begin{equation}
|\Gamma_\sigma \Gamma_n\rho\Gamma_\omega \kappa (LS) J M_{J}\rangle 
{\textstyle 
= \left[{\mathcal{P}^{\Gamma_n}(A^{(2\,0)})
\times 
|\Gamma_\sigma}\rangle\right]^{\rho\Gamma_\omega}_{\kappa 
(LS) J M_{J}},
}
\label{bs}
\end{equation}
where $\Gamma_\sigma$ $\equiv $ $N_\sigma\left(\lambda_{\sigma}\, \mu_{\sigma}\right)$
labels \SpR{3} irreps with
$\left(\lambda_{\sigma}\,\mu_{\sigma}\right)$ denoting a \SU{3}
lowest-weight state,
$\Gamma_{n}\equiv n\left(\lambda_{n}\, \mu_{n}\right)$, and
$\Gamma_\omega\equiv N_\omega\left(\lambda_{\omega}\,
\mu_{\omega}\right)$.
The $\left(\lambda_{n}\, \mu_{n}\right)$ set
gives the
overall \SU{3} symmetry of
$\frac{n}{2}$ coupled raising 
operators in $\mathcal{P}$, $\left(\lambda_{\omega}\,
\mu_{\omega}\right)$ specifies the \SU{3} symmetry of the
symplectic state, and $N_\omega=N_\sigma+n$ is the total number of
oscillator quanta related to the eigenvalue,
$N_\omega \hbar\Omega$, of a HO Hamiltonian that is free of 
spurious modes.

The symplectic raising operator $A^{(2\,0)}_{lm}$, 
which is a \SU{3} tensor 
with $\left(\lambda\, \mu\right)=\left(2\,0\right)$  character,
can be expressed as a bilinear product of the HO raising operators,
\begin{equation}
{\textstyle
A^{(20)}_{lm}= \frac{1}{\sqrt{2}} \sum_{i}
\left[b_{i}^{\dagger}\times b_{i}^{\dagger}\right]^{(20)}_{lm} 
- \frac{1}{\sqrt{2}A} \sum_{s,t}
\left[b^{\dagger}_{s}\times b^{\dagger}_{t}\right]^{(20)}_{lm},
}
\label{A20}
\end{equation}
where the sums are over all $A$ particles of the system. The first term in (\ref{A20})
describes 2\ho~ one-particle-one-hole (1p-1h) excitations (one particle raised by two shells) and the
second term eliminates the spurious center-of-mass excitations in the construction (\ref{bs}). 
For the purpose of comparison to NCSM results, the basis states of the 
$\ket{\Gamma_{\sigma}}$ bandhead
in (\ref{bs}) are constructed in a $m$-scheme basis,
\begin{eqnarray}
&&\ket{\Gamma_{\sigma}\kappa (L_0S_0) J_0 M_{0}} {\textstyle =} \nonumber \\
&&\left[\mathcal{P}^{(\lambda_{\pi}\,\mu_{\pi})}_{S_{\pi}}(a^{\dagger}_{\pi})
{\textstyle \times}
\mathcal{P}^{(\lambda_{\nu}\,\mu_{\nu})}_{S_{\nu}}(a^{\dagger}_{\nu})
\right]^{(\lambda_{\sigma}\,\mu_{\sigma})}_{\kappa (L_0S_0) J_0 M_{0}}\ket{0},
\label{bandhead}
\end{eqnarray}
where \ket{0} is a vacuum state,
$\mathcal{P}^{(\lambda_{\pi}\,\mu_{\pi})}_{S_{\pi}}$ and
$\mathcal{P}^{(\lambda_{\nu}\,\mu_{\nu})}_{S_{\nu}}$ denote polynomials of
proton ($a^{\dagger}_{\pi}$) and neutron ($a^{\dagger}_{\nu}$) creation
operators coupled to good
\SU{3}$\times$\SU{2} symmetry.

\section{Results and Discussions}
The lowest-lying states of $^{12}$C and $^{16}$O were calculated using the
NCSM as implemented through the Many Fermion Dynamics ~(MFD) code \cite{Vary92_MFD}.
For $^{12}$C we used an effective interaction derived from the realistic JISP16
{\it NN} potential \cite{ShirokovMZVW04} for different \ho~ oscillator strengths,
while for $^{16}$O the bare JISP16 interaction was employed.
We are particularly interested in the $J\!=\!0^{+}_{gs}$ and the
lowest $J\!=\!2^{+}(\equiv\!2^{+}_{1})$ and $J\!=\!4^{+}(\equiv\!4^{+}_{1})$ states
of the ground-state (gs) rotational band in $^{12}$C and the $J\!=\!0^{+}_{gs}$ state
in $^{16}$O that appear to be well converged  in the $N_{max}=6$ NCSM basis space.

Here we report on an  analysis that is restricted to \ph{0} configurations. It is
important to note that 2\ho~\ph{2} (2 particles raised by one  shell each) and higher
rank \ph{n} excitations and allowed multiples thereof can be included by building them
into an expanded set of lowest-weight
\SpR{3} starting state configurations. The same ``build-up'' logic, (\ref{bs}), holds
because by construction these additional starting state configurations are also
required to be lowest-weight \SpR{3} states.  Note that if one were to include all
possible lowest-weight \ph{n} starting state configurations $(n \leq N_{max})$, and
allowed multiples thereof, one would span the entire NCSM space. The addition of
2\ho~ \ph{2}, 4\ho~ \ph{4}, and higher configurations,  which build upon more complex
starting states, will be the subject of a follow-on investigation.

\subsection{Ground-state rotational band in the $^{12}$C nucleus}
For $^{12}$C there are $13$ unique \ph{0} \SpR{3} irreps which form the
symplectic bandhead basis states, $\ket{\Gamma_{\sigma}}$ with $N_\sigma =24.5$. For
each \ph{0} \SpR{3} irrep we generated basis states according to (\ref{bs})
up to $N_{max}=6$ (6\ho~ model space).
The typical dimension of a symplectic irrep basis in
the $N_{max}=6$ space is on the order of $10^{2}$ as compared to
$10^{7}$ for the full NCSM $m$-scheme basis space.

\begin{figure}[b]
\centerline{\epsfig{file=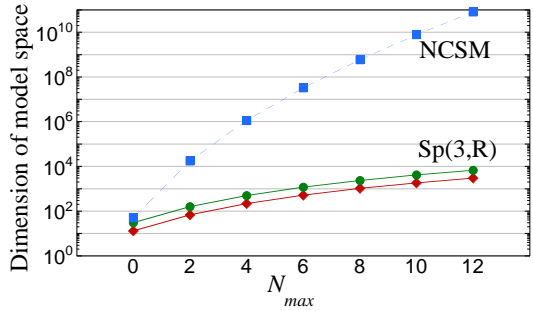,width=70mm}}
\caption{Dimension of the NCSM (blue squares) and $J=0,2,$ and $4$ \SpR{3} 
(red diamonds for the 3 most significant \ph{0} irrep case and green
circles for when all 13 \ph{0} irreps are included) model spaces as a function of
maximum allowed \ho~ excitations, $N_{max}$.\label{dimMdlSpace}}
\end{figure}

As $N_{max}$ is increased the dimension of the $J=0,2,$ and
$4$ symplectic space  built on the \ph{0} \SpR{3} irreps
grows very slowly compared to the NCSM space dimension (Fig. \ref{dimMdlSpace}). This means that a
space spanned by a set of symplectic basis states may be computationally
manageable
even when high-\ho~ configurations are included.

Analysis of overlaps of the symplectic states with the NCSM eigenstates
for the $0^+_{gs}$ and the lowest $2^+$ and $4^+$ states reveals
nonnegligible overlaps for
only 3 of the 13 \ph{0} \SpR{3} ($N_\sigma =24.5$) irreps, specifically, the
leading (most deformed) representation $\left(\lambda_{\sigma}~\mu_{\sigma}\right)=(0~4)$
carrying spin $S=0$ together with two
$S=1$ representations with identical labels (1 2) but different bandhead constructions for
protons and neutrons (\ref{bandhead}), namely, 
$\{(\lambda_{\pi}\,\mu_{\pi})S_{\pi},\
(\lambda_{\nu}\,\mu_{\nu})S_{\nu}\}$ is $\{(0\;2)0,\ (1\;0)1\}$ and $\{(1\;0)1,\ (0\;2)0\}$.
The dominance of
only three irreps additionally reduces the dimensionality of the symplectic model
space (Fig. \ref{dimMdlSpace}, red diamonds).

The overlaps of the most dominant symplectic states with the NCSM eigenstates for the
$0^+_{gs}$, $2^+_1$ and $4^+_1$ states in the $0$, $2$, $4$
and $6\hbar\Omega$ subspaces are given in Table~\ref{TABLE_15MeV}. The
results show that approximately 80\% of the NCSM eigenstates fall within
a subspace spanned by the 3 leading \ph{0} \SpR{3} irreps, with the
most deformed irrep, $(0~4)$, carrying about 65\% of the 80\%. 
In order to speed up the calculations, we retained only the largest amplitudes
of the NCSM states, those sufficient to account for at least 98\% of the norm which
is quoted also in the table.
\begin{table} [hb]
\caption{Probability distribution of NCSM eigenstates for $^{12}$C across the leading 3 \ph{0}
\SpR{3} irreps, \ho=15 MeV.\label{TABLE_15MeV}}
\smallskip
\begin{small}\centering
\begin{tabular*}{\textwidth}{@{\extracolsep{\fill}}crrrrrrrrr}
\hline  \noalign {\smallskip}
& & $0\hbar\Omega$ & $2\hbar\Omega$ & $4\hbar\Omega$ & $6\hbar\Omega$ & Total \\
\hline
\multicolumn{7}{c}{J=0} \\
\hline
              & $(0\;4)S=0$ & $46.26$ & $12.58$ & $4.76$ & $1.24$ & $64.84$\\
   $\SpR{3}$  & $(1\;2)S=1$ & $4.80$ & $2.02$ & $0.92$ &  $0.38$  & $8.12$\\
  	      & $(1\;2)S=1$ & $4.72$ & $1.99$ & $0.91$ &  $0.37$  & $7.99$ \\
\cline{2-7}
             & Total  & $55.78$ & $16.59$ & $6.59$ &  $1.99$  & $80.95$ \\
 NCSM        &     & $56.18$ & $22.40$ & $12.81$ &  $7.00$  & $98.38$ \\
\hline
\multicolumn{7}{c}{J=2} \\
\hline
    & $(0\;4)S=0$ & $46.80$ & $12.41$ & $4.55$ & $1.19$ & $64.95$\\
 $\SpR{3}$ & $(1\;2)S=1$ & $4.84$ & $1.77$ & $0.78$ &  $0.30$  & $7.69$ \\
& $(1\;2)S=1$ & $4.69$ & $1.72$ & $0.76$ &  $0.30$  & $7.47$ \\
\cline{2-7}
    &    Total     & $56.33$ & $15.90$ & $6.09$ &  $1.79$  & $80.11$ \\
  NCSM  &           & $56.63$ & $21.79$ & $12.73$ &  $7.28$ & $98.43$ \\
\hline
\multicolumn{7}{c}{J=4} \\
\hline
    & $(0\;4)S=0$ & $51.45$ & $12.11$ & $4.18$ & $1.04$  & $68.78$\\
  $\SpR{3}$ & $(1\;2)S=1$ & $3.04$ & $ 0.95$ & $0.40$ &  $0.15$   & $4.54$ \\
  & $(1\;2)S=1$ & $3.01$ & $0.94$ & $0.39$ &  $0.15$   & $4.49$ \\
\cline{2-7}
    &    Total & $57.50$ & $14.00$ & $4.97$ &  $1.34$    & $77.81$ \\
  NCSM  &            & $57.64$ & $20.34$ & $12.59$ &  $7.66$     & $98.23$ \\
\hline
\end{tabular*}
\end{small}
\end{table}

In addition, the $0^{+}_{gs}$ analysis of the $S=0$ ($S=1$) part of the NCSM
wavefunction  reveals that within each \ho~ subspace only about  $1-1.5\%$ of the NCSM
$0^{+}_{gs}$ are not accounted for by the $S=0$ ($S=1$) \SpR{3} irrep(s) under
consideration. In the $N_{max}=6$ model space the
$S=0$ symplectic irrep and the two $S=1$ irreps account for 91\% and 80\%,
respectively,  of the corresponding
$S=0$ and $S=1$ parts of the NCSM realistic eigenstate for the $J\!=\!0^{+}_{gs}$ in
$^{12}$C. In summary, the $S=0$ plus $S=1$ part of the NCSM wavefunction is very well
explained by only the three \SpR{3} collective configurations.

\begin{figure}[t]
\centerline{\epsfig{file=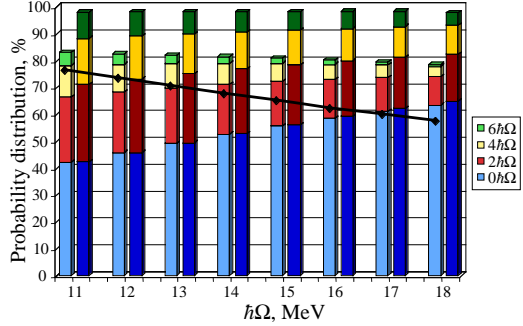,width=70mm}}
\caption{$^{12}$C ground $0^+$ state probability distribution over $0$\ho~ (blue,
lowest) to
$6$\ho~ (green, highest) subspaces for the 3 \ph{0} \SpR{3} irrep case (left) and NCSM
(right) together with the $(0~4)$ irrep contribution (black diamonds) as a function of
the \ho~ oscillator strength in MeV for
$N_{max}=6$.\label{C12prblty_vs_hw}}
\end{figure}

How the results presented in Table~\ref{TABLE_15MeV} change as a function of
the oscillator strength \ho~ is shown in Fig. \ref{C12prblty_vs_hw} for the case of
the $0^+_{gs}$ state. Clearly, the projection of the NCSM wavefunctions onto the
symplectic space slightly changes as one varies the oscillator strength \ho. The 3
\SpR{3} irreps, (0~4)$S=0$ and the two (1~2)$S=1$, remain dominant, only their
contributions change. The overall overlaps increase towards smaller \ho~HO frequencies
and, for example, for
$0^+_{gs}$ it is 85\% in the $N_{max}=6$ and $\ho=11$MeV case. Clearly, the largest
contribution comes from the leading, most deformed, $(0~4)S=0$ \SpR{3} irrep, growing
to 91\% of the total \SpR{3}-symmetric part for \ho~=11 MeV. As expected, Fig.
\ref{C12prblty_vs_hw} also confirms that with increasing $\hbar\Omega$ the higher \ho~
excitations contribute less while the lower 0\ho~ configurations grow in importance. 

In short, the low-lying states in $^{12}$C are well described in terms of only three
\SpR{3} irreps with total dimensionality of 514, which is only 0.001\% of the NCSM
space, with a clear dominance of the most deformed
$(0\,4)S=0$ collective configuration. It is important to note that our 
results suggest  that overlaps can
be further improved by the inclusion of the most  important 2\ho~2p-2h 
\SpR{3} irreps. In this way it may be
possible to achieve overlaps of more then 90\%  while keeping the size of the basis space small, 
possibly much less than ~1\% of the NCSM  result. This is the subject of ongoing investigations and
will be addressed in  a subsequent study.

The $0^{+}_{gs}$, $2^{+}_1$ and $4^{+}_1$ states, constructed in terms of the
three \SpR{3} irreps with probability  amplitudes defined by the
overlaps with the NCSM wavefunctions, were also used to determine
B(E2) transition rates. The $B(E2:2^+_1 \rightarrow 0^+_{gs})$ value,
for example, turns out to be $\approx$110\% of the corresponding NCSM
number for the $\hbar\Omega=15$MeV and $N_{max}=4$ case. While this ratio decreases
slightly for smaller $\hbar\Omega$ oscillator strengths, it is
significant that this estimate for the dominant \SpR{3} configurations
exceeds the corresponding full NCSM results and therefore lies closer
to the experimental $B(E2:2^+_1 \rightarrow 0^+_{gs})$ value. 

\subsection{Ground state in the $^{16}$O nucleus}

\begin{figure}[t]
\centerline{\epsfig{file=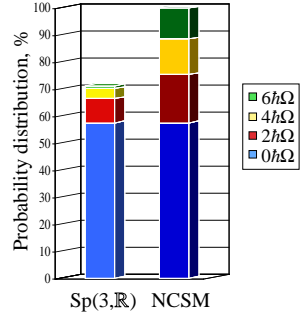,width=40mm}}
\caption{$^{16}$O ground $0^+$ state probability distribution  over $0$\ho~ (blue,
lowest) to $6$\ho~ (green, highest) subspaces for the leading \ph{0} (0~0) \SpR{3}
irrep case (left) and NCSM (right) for $N_{max}=6$ and bare JISP16
interaction.\label{O16prblty}}
\end{figure}

The Sp-NCSM is also applied to the ground state of a closed-shell nucleus like
$^{16}$O. There is only one \ph{0} \SpR{3} irrep with spin $S=0$ and $\Gamma_{\sigma}$
specified by $N_\sigma =34.5$ and $(\lambda_\sigma~\mu_\sigma)=(0~0)$. As in the
$^{12}$C case, for the \ph{0} \SpR{3} irrep we generated basis states according to
(\ref{bs}) up to $N_{max}=6$ (6\ho~ model space), which yields a symplectic model space
that is only a fraction ($\approx 0.1\%$) of the size of the NCSM space. Consistent
with the outcome for $^{12}$C, the projection of the NCSM eigenstates onto the
symplectic basis reveals a large
\SpR{3}-symmetric content in the ground-state wavefunction (Fig.
\ref{O16prblty}). Furthermore, the overall overlap  increases by $\approx 10\%$ when
the most significant 2\ho~ \ph{2} are included.
\newline

While the focus here has been on demonstrating the existence of
\SpR{3} symmetry in NCSM results for $^{12}$C and $^{16}$O, and therefore a
possible path forward for extending the NCSM to a Sp-NCSM scheme, the
results can also be interpreted as a further strong confirmation of Elliott's \SU{3}
model since the projection of the NCSM states onto the 0$\hbar\Omega$ space [Fig.
\ref{C12prblty_vs_hw} and Fig.\ref{O16prblty}, blue (right) bars] is a
projection of the NCSM results onto the \SU{3} shell model. For  $^{16}$O the
0$\hbar\Omega$ \SU{3} symmetry is $\approx 60\%$ of the NCSM $0^+_{gs}$ [Fig.
\ref{O16prblty}, blue (left) bars]. For $^{12}$C the 0$\hbar\Omega$ \SU{3} symmetry 
ranges from just over 40\% of the NCSM $0^+_{gs}$ for
$\hbar\Omega$ = 11 MeV to nearly 65\% for
$\hbar\Omega$ =18 MeV [Fig. \ref{C12prblty_vs_hw}, blue (left) bars] with 80\%-90\% of this symmetry
governed by the leading (0\,4) irrep.  These numbers are consistent with what has been shown to
be a dominance of the leading \SU{3} symmetry for \SU{3}-based shell-model studies with
realistic interactions in 0$\hbar\Omega$ model spaces.  It seems the simplest of
Elliott's collective states can be regarded as a good first-order approximation in the
presence of realistic interactions, whether the latter is restricted to a
0$\hbar\Omega$ model space or the richer multi-$\hbar\Omega$ NCSM model
spaces.

\section{Conclusions}
Wavefunctions, which are  obtained in a NCSM analysis with the
JISP16 realistic interaction, project at approximately the 80\% level onto the
leading (three) \ph{0} irreps of the corresponding Sp-NCSM for the lowest $0^+_{gs}$,
$2^+_1$ and $4^+_1$ states in $^{12}$C and at more than 70\% level
for the ground state in the closed-shell $^{16}$O nucleus. (While not part of the
current analysis, preliminary results indicate that when the space is expanded to
include the most important 2$\hbar\Omega$
\ph{2} irreps the percentage grows by approximately 10\%.) The results
confirm for the first time the validity of the \SpR{3} approach when
realistic interactions are invoked and hence demonstrate the
importance of the \SpR{3} symmetry in light nuclei as well as
reaffirm the value of the simpler \SU{3} model upon which it is based. 

The results further suggest that a Sp-NCSM extension of the NCSM may be a
practical scheme for achieving convergence to measured B(E2) values
without the need for introducing an effective charge and even
for modeling cluster-like phenomena as these modes can be
accommodated within the general framework of the \SpR{3} model if
extended to large model spaces (high $N_{max}$), but with a size that is typically only
a fraction of the NCSM size. This suggests that a Sp-NCSM code could allow one to
extend  no-core calculations to higher \ho~ configurations and heavier nuclei that are 
currently unreachable because the model space is typically too large to 
handle, even on the best of modern day computers. 

\section*{Acknowledgments}

This work was supported by the US National Science Foundation, Grant Nos 0140300 \&
0500291, and the Southeastern Universities Research Association, as well as, in part,
by the US Department of Energy Grant Nos.  DE-AC02-76SF00515 and DE-FG02-87ER40371 and
at the University of California, Lawrence Livermore National Laboratory under contract
No. W-7405-Eng-48. Tom\'a\v{s} Dytrych acknowledges supplemental support from the
Graduate School of Louisiana State University.

\end{document}